\documentclass{article}
\usepackage{arxiv}
\usepackage{graphicx}
\usepackage{natbib}
\usepackage{hyperref}
\usepackage{listings}
\usepackage{xcolor}
\usepackage[onehalfspacing]{setspace}
\usepackage{multirow}
\usepackage{colortbl}
\usepackage{rotating}

\begin{document}

\title{A Review of Media Copyright Management using Blockchain Technologies from the Academic and Business Perspectives}


\author{ \href{https://orcid.org/0000-0003-2207-9605}{\includegraphics[scale=0.06]{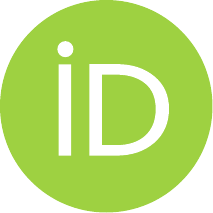}\hspace{1mm}Roberto García}\thanks{Corresponding author} \\
	Computer Science and Industrial Engineering\\
	Universitat de Lleida\\
	Víctor Siurana, 1, 25003 Lleida, Spain \\
	\texttt{roberto.garcia@udl.cat} \\
	\And
	\href{https://orcid.org/0000-0002-8495-2505}{\includegraphics[scale=0.06]{orcid.pdf}\hspace{1mm}Ana Cediel} \\
	Department of Public Law\\
	Universitat de Lleida\\
	Víctor Siurana, 1, 25003 Lleida, Spain \\
	\texttt{anna.cs@udl.cat} \\
	\And
	\href{https://orcid.org/0000-0001-9124-1464}{\includegraphics[scale=0.06]{orcid.pdf}\hspace{1mm}Mercè Teixidó} \\
	Computer Science and Industrial Engineering\\
	Universitat de Lleida\\
	Víctor Siurana, 1, 25003 Lleida, Spain \\
	\texttt{merce.teixido@udl.cat} \\
	\And
	\href{https://orcid.org/0000-0001-6304-9635}{\includegraphics[scale=0.06]{orcid.pdf}\hspace{1mm}Rosa Gil} \\
	Computer Science and Industrial Engineering\\
	Universitat de Lleida\\
	Víctor Siurana, 1, 25003 Lleida, Spain \\
	\texttt{rosamaria.gil@udl.cat} \\
}


\renewcommand{\shorttitle}{A Review of Media Copyright Management using Blockchain Technologies}

\hypersetup{
pdftitle={A Review of Media Copyright Management using Blockchain Technologies from the Academic and Business Perspectives},
pdfsubject={q-cs.NC, q-bio.QM},
pdfauthor={Roberto García, Ana Cediel, Mercè Teixidó, Rosa Gil},
pdfkeywords={copyright, media, blockchain, digital rights management, social media, business, review},
}

\maketitle

\begin{abstract}
 Blockchain technologies open new opportunities for media copyright management. To provide an overview of the main initiatives in this blockchain application area, we have first reviewed the existing academic literature. The review shows literature is still scarce and immature in many aspects, which is more evident when comparing it to initiatives coming from the industry. Blockchain has been receiving significant inflows of venture capital and crowdfunding, which have boosted its progress in many fields, including its application to media management. Consequently, we have complemented the review with a business perspective. Existing reports about blockchain and media have been studied and consolidated into four prominent use cases. Moreover, each one has been illustrated through existing businesses already exploring them. Combining the academic and industry perspectives, we provide a more general and complete overview of current trends in media copyright management using blockchain technologies.
\end{abstract}

\keywords{copyright, media, blockchain, digital rights management, social media, business, review}

\section{Introduction}

Blockchain technologies have opened new opportunities for media copyright management after previous shifts caused by digitization or communication networks \cite{Serrao2010}. In some cases, blockchain promises solutions to problems resulting from those previous shifts like the ease of copying or uncontrolled digital distribution \cite{Li2020}. 

We aim to provide an overview of recent contributions addressing media copyright management using blockchain technologies. Considering just an academic perspective, our contribution goes beyond the state of the art as it is the first review paper about blockchain for copyright management, as the literature overview in Section~\ref{Sec:LiteratureOverview} shows.

Moreover, the contribution goes beyond just analyzing the topic from an academic perspective and shows that reducing the study to just that point of view is not enough to build a clear picture of the domain. On the contrary, our results show that it is crucial to also consider the business perspective as it is where most of the advancements regarding the use of blockchain for copyright management are being generated. By complementing the academic with the business perspective, it should be possible to provide a more complete overview of the most relevant contributions and trends. 

To summarise the aim of this review, these are the research questions being addressed:

\begin{itemize}
    \item \textbf{RQ1}: is the application of blockchain technologies to media copyright management a mature academic research area?
    \item \textbf{RQ2}: which are the main areas of academic research dealing with media copyright management using blockchain?
    \item \textbf{RQ3}: which are the main business use cases for media copyright management using blockchain?
    \item \textbf{RQ4}: where do most "blockchain for media copyright management" originate, academia or industry?
\end{itemize}

The rest of the paper is organized as follows. Next, Section~\ref{Sec:IntroCopyrightBlockchain} presents copyright management and the contributions it can receive from blockchain technologies. Then, Section~\ref{Sec:LiteratureReview} overviews the state of the art of academic research through an analysis of the Scopus database and tries to answer to research questions RQ1 and RQ2. We complete the review with the business perspective in Section~\ref{Sec:BusinessReview}, where the main business cases and examples of initiatives in each of them are presented, addressing RQ3 and RQ4. Finally, Section~\ref{Sec:Discussion} presents the conclusions regarding the research questions drawn from these reviews, from both the academic and business points of view.

\subsection{Copyright Management and Blockchain} \label{Sec:IntroCopyrightBlockchain}

\begin{figure}[ht]
\centerline{\includegraphics[width=0.8\textwidth]{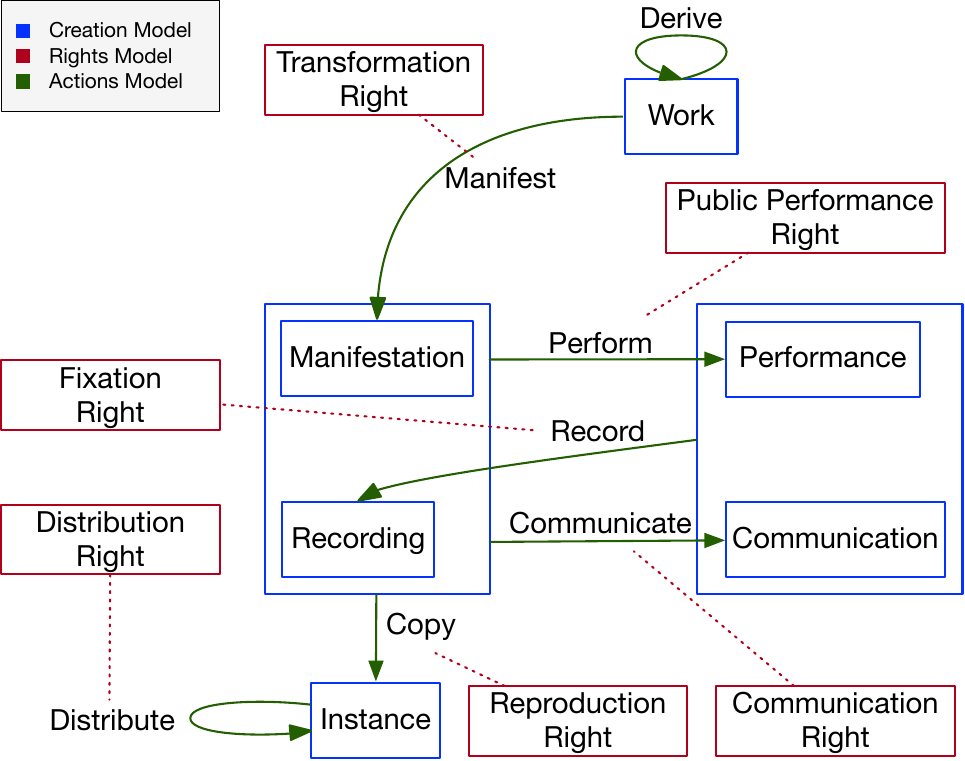}}
\caption{ The copyright life cycle as represented by the Copyright Ontology \cite{Garcia2010}. From \cite{CopyrightOnto2022} with permission}
\label{Fig:CopyrightOntology}
\end{figure}

The full copyright lifecycle, from its generation, when a creator first manifests a new work to its consumption through different embodiments, from physical or digital objects to performances or media streams \cite{Garcia2010}. The copyright lifecycle view provided by the Copyright Ontology summarises it as shown in \textbf{Figure~\ref{Fig:CopyrightOntology}}. The ontology model includes the different "stages" creations can go through (Creation Model), the actions that move creations along their lifecycle (Actions Model) and the rights that restrict these actions (Rights Model).

The first step in the copyright value chain is when the creator embodies the creation into something tangible (a Manifestation). That manifestation can be used to claim authorship if it is the first time the underlying work (the abstract idea behind the creation) has been manifested.

The way to decide who the original creator is in case of dispute is to determine who first manifested the creation. Then, the other creators might have had access to it and just made an unoriginal copy. Alternatively, it might be considered a derivation if it is not an exact copy and sufficiently original. In that case, manifesting this derivation is regulated by the Transformation Right.

The motivation to use blockchain technologies to support this part of the copyright life cycle is because they facilitate time-stamping those manifestations and linking them to the claimed creator in a decentralized and trustless way, i.e., that does not require trusted third parties and centralized registries.

From this initial step of setting authorship and associating all copyright to the original creator, the whole copyright value chain emerges, regulated by different rights, through actions like performing a creation (e.g. a music composition), recording or streaming it.

Blockchain also contributes by transparently tracking all these actions along the copyright value chain, facilitating splitting royalties’ payments to all the involved actors (for instance: composer, performer, lyricist, label, etc.) using smart contracts in a more timely manner than current systems, where artists might need to wait years to receive their payments.

Additionally, blockchain can control rights themselves, bookkeeping who owns the different kinds of rights on a particular creation, including their temporal and territorial dimensions. This control includes who holds the rights, the percentage held, tracking rights transfers, calculating royalties’ splits based on those rights, etc.

\section{Literature Review} \label{Sec:LiteratureReview}

The literature review explores existing academic publications addressing media copyright management using blockchain technologies. It starts with an overview of the literature based on bibliometric analysis, in Section~\ref{Sec:LiteratureOverview}, and then conducts a more detailed literature review by first clustering the papers based on their content and then studying some of the most representative publications per cluster in Section~\ref{Sec:LiteratureAnalysis}.

The analysis is based on statistical and visualization tools applied to the set of papers resulting from queries to academic literature databases, concretely those provided by Scopus to analyse query results. We have used the Scopus database, which includes both high-quality journals but, contrary to other databases like Web of Science, it also includes conferences, where most of the research about blockchain is currently being published as shown later. The query to retrieve the relevant documents about media copyright management using blockchain from Scopus is shown in \textbf{Table \ref{Tab:ScopusQuery}}.

\begin{table}[!ht]
\centering
\begin{tabular}{|c|}
\hline
\\
( TITLE(right OR copyright) OR KEY(right OR copyright) ) \\
AND TITLE-ABS-KEY ( media AND blockchain ) \\
AND ( LIMIT-TO ( LANGUAGE, "English" ) ) \\
\\
\hline
\end{tabular}
\caption{Scopus query for media copyright management and blockchain documents.}
\label{Tab:ScopusQuery}
\end{table}

The query is more complex than expected because the "right" or "copyright" terms are common in publication abstracts or as part of the paper text, even when the paper has nothing to do with these topics. They appear in the abstract or at the end of the paper body as part of the typical copyright statements added by publishers, e.g. "(c) Copyright 2020" or "all rights reserved".

This fact introduces a lot of noise in the results, going from almost 2.000 results, if we look for "copyright" or "right" in the abstract, to 38 using the final version of the query on October 20th 2022, which only looks for "right" or "copyright" in the title or the keywords of documents in English.

We used an equivalent search with the Web of Science database. However, it produced less than half of the Scopus results, and all the relevant ones were already in Scopus. It is also important to note that the query is for any publication without restricting it to a predefined time span.

Next, we provide an overview of the Scopus' results using different points of view (publication years, subject areas and publication types) and then perform a more detailed analysis based on their content.

\subsection{An Overview}\label{Sec:LiteratureOverview}

\begin{figure}[htb]
\centerline{\includegraphics[width=0.8\columnwidth]{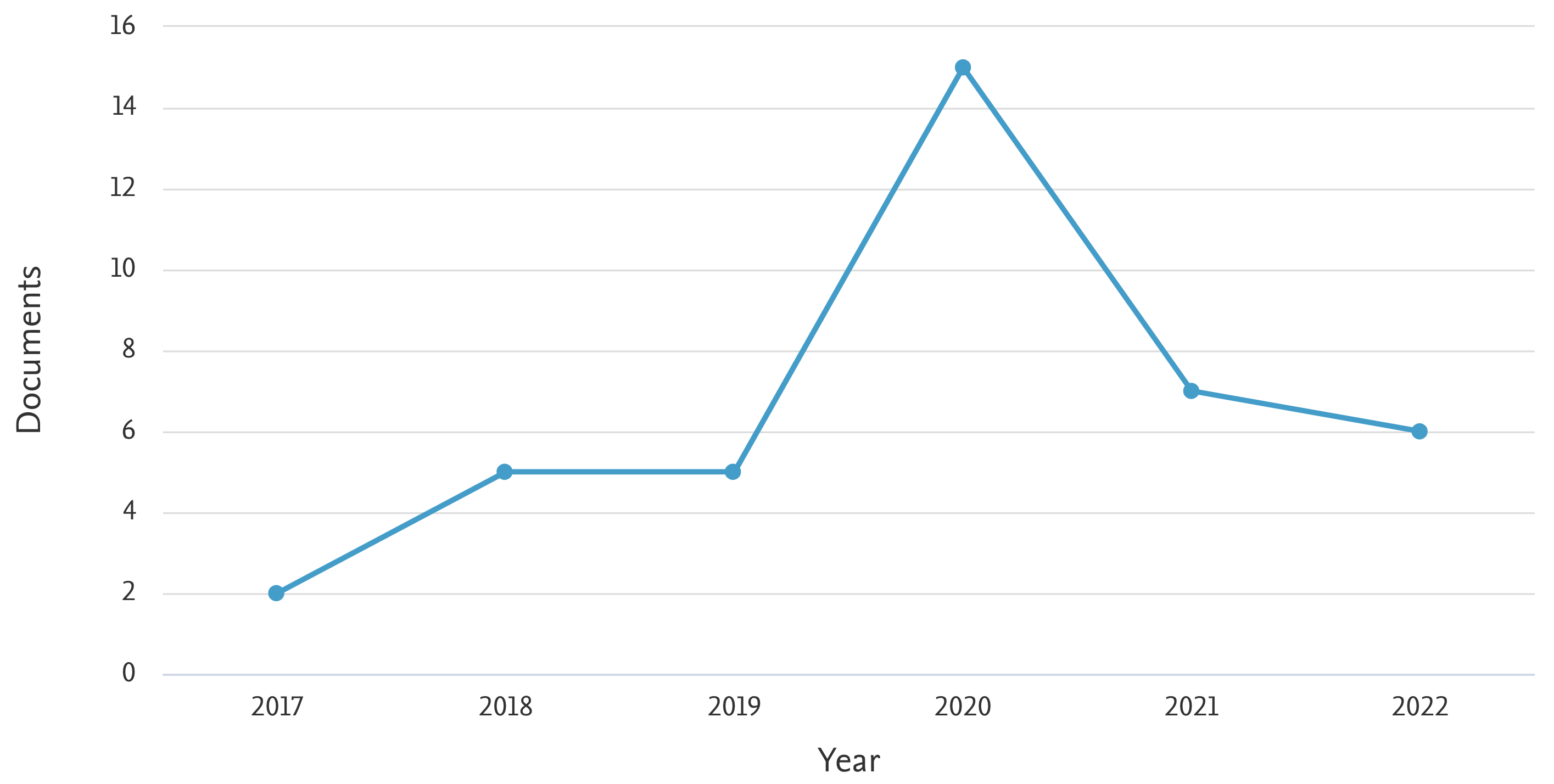}}
\caption{Number of publications per year in Scopus about media copyright and blockchain}
\label{Fig:ScopusPerYear}
\end{figure}

Blockchain-based copyright management is a very young topic and, as shown in \textbf{Figure~\ref{Fig:ScopusPerYear}}, we have been able to retrieve publications based on Table~\ref{Tab:ScopusQuery} starting from 2017. It is also important to note that despite the significant increase in outputs during 2020, moving from 5 publications in 2019 to 15, the results for 2021 and 2022 are back down to 7 and 6 papers respectively. Though it is too early to draw long-term conclusions, it seems that for the moment, this is not a topic drawing a lot of attention from the academic research community. Especially, as we will see later in Section~\ref{Sec:BusinessReview}, if we compare it to the amount of activity in the business domain.

\begin{figure}[htb]
\centerline{\includegraphics[width=0.8\columnwidth]{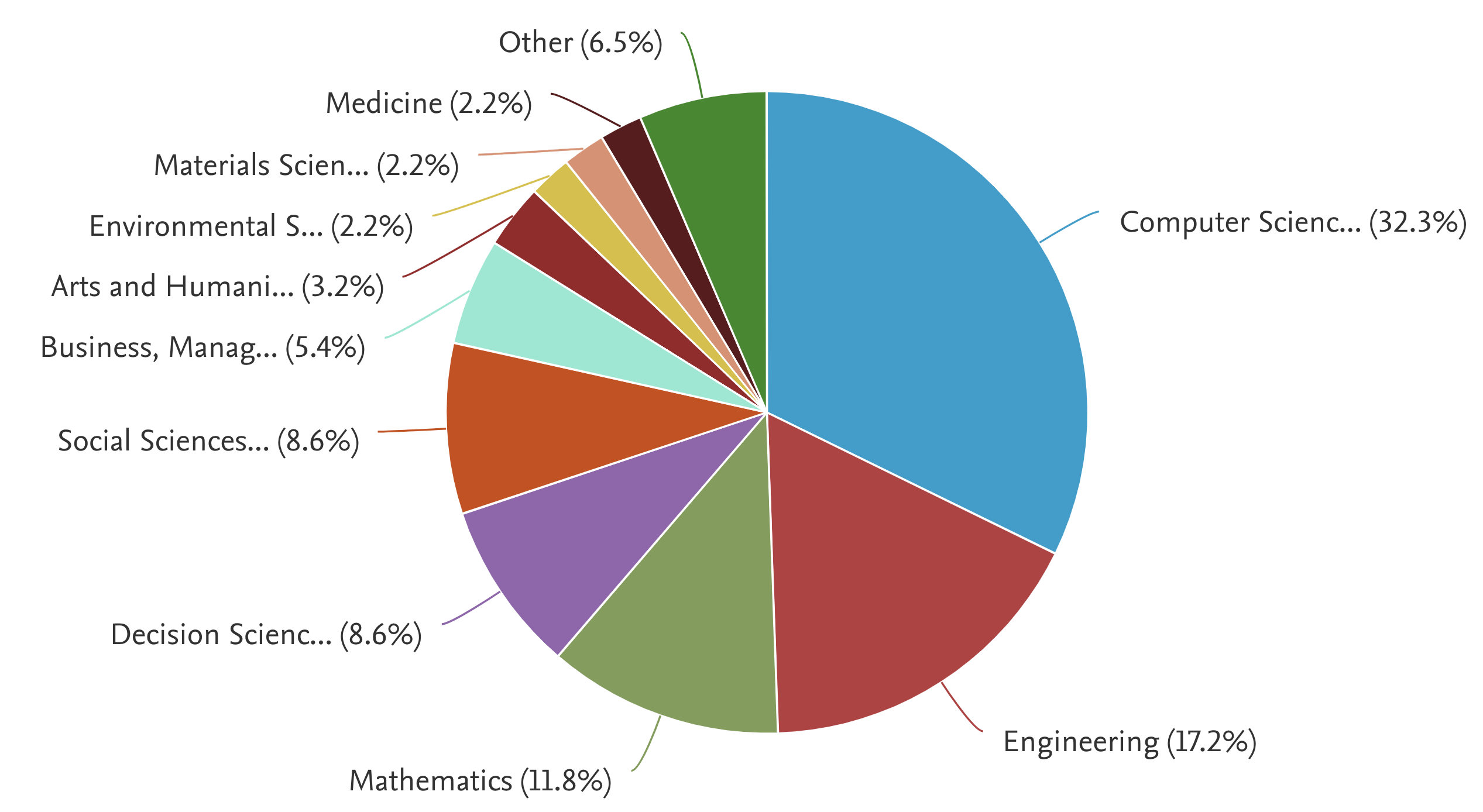}}
\caption{Main subject areas of the publications in Scopus for the query about media copyright and blockchain}
\label{Fig:ScopusPerSubject}
\end{figure}

The documents retrieved from Scopus can be grouped by different subject areas as shown in \textbf{Figure~\ref{Fig:ScopusPerSubject}}. The three most common subject areas are Computer Science (32.3\%), Engineering (17.2\%) and Mathematics (11.8\%). These three subject areas alone account for more than 60\% of the retrieved literature, showing that most focus is on the technological foundations of blockchain applied to media copyright. On the other hand, contributions in other areas like Social Sciences or Business and Management are still scarce, representing 8.6\% and 5.4\% respectively.

\begin{figure}[htb]
\centerline{\includegraphics[width=0.8\columnwidth]{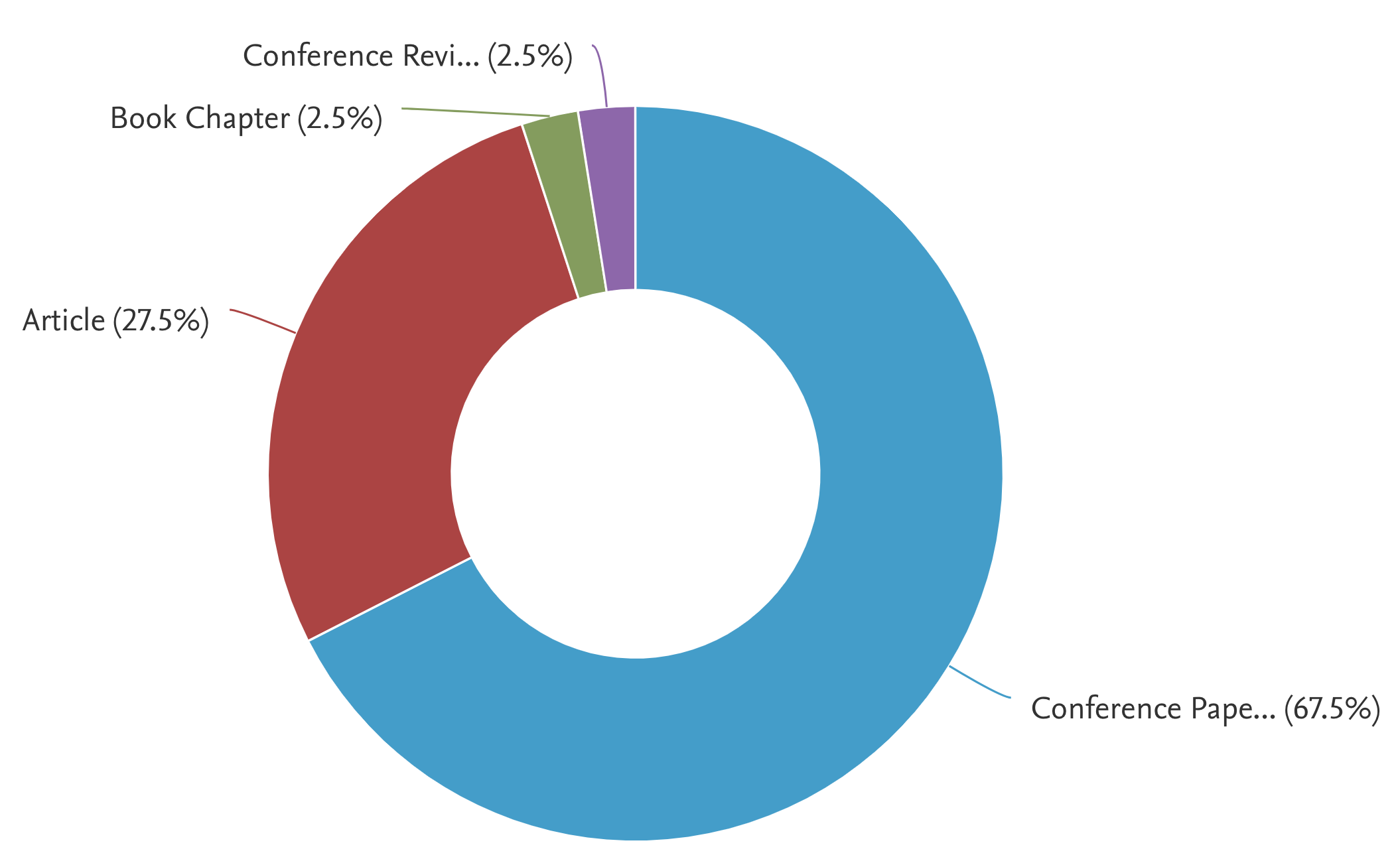}}
\caption{Document types for the Scopus query about media copyright and blockchain}
\label{FigScopusPerType}
\end{figure}

To complete this literature overview, \textbf{Figure~\ref{FigScopusPerType}} shows results based on the type of document. The most common one is the Conference Paper accounting for slightly more than two-thirds of the documents. The other third is mainly articles in journals. The prevalence of documents in conferences usually signals that most research is still in the early stages \cite{Kim2019}. In this case, the publication ratio between the number of conference papers minus the journal ones divided by the total amount of documents is 0.4. For comparison, the average ratio in Computer Science, the discipline with a higher ratio of documents published in conferences, is 0.15, being a ratio of 1 complete dominance of conferences. 

Another indicator of the lack of maturity of this research area in academia is that there is just one review document, and it is a conference review providing an overview of just the proceedings of a conference \cite{Katsikas2020}. Moreover, the conference does not include any of the selected documents, just one about social media and another about e-voting using blockchain that, combined, made the query match the conference review.

\subsection{Analysis of the Relevant Literature} \label{Sec:LiteratureAnalysis}

We have partially automated the detailed analysis of the literature using Bibliometrix \cite{Aria2017}, which has received as input the selected 37 documents after excluding the conference review paper discarded in the previous section. This tool makes it possible to cluster the analyzed documents based on their abstract and keywords as detailed by \cite{Aria_Corrado2022}. Following this approach, we have identified four topics that we can use to categorize the literature about media copyright management using blockchain: \textit{Digital Rights Management}, \textit{Copyright Protection}, \textit{Social Media}, and \textit{Intellectual Property Rights}. Next, we detail each of these topics and the documents corresponding to each of them are listed in \textbf{Table~\ref{Tab:ScopusClusters}}.

\begin{table}[htb]
\begin{tabular}{lp{0.68\textwidth}}
\textbf{Main Topics} & \textbf{Documents} \\
\hline
\\
Digital Rights Management & \cite{HOLLAND2017}, \cite{XU2017}, \cite{CHEN2018}, \cite{ENGELMANN2018}, \cite{HOLLAND2018}, \cite{KUO2019}, \cite{XIN2020}, \cite{SCHNEIDER2020}, \cite{GARBA2021}, \cite{GARCIA2021}, \cite{Ramani2022}, \cite{Geethanjali2022} \\
Copyright Protection & \cite{BHOWMIK2018}, \cite{QURESHI2019}, \cite{AGYEKUM2019}, \cite{ZHAO2020}, \cite{DOBRE2020}, \cite{JIANG2020}, \cite{YANG2020}, \cite{TEMMERMANS2020}, \cite{CUI2021}, \cite{Igarashi2021}, \cite{Stallin2022}, \cite{Yang2022a}, \cite{Yang2022b} \\
Social Media & \cite{TRIPATHI2019}, \cite{GARCIA2019}, \cite{DASKAL2020}, \cite{MILKOVIC2020}, \cite{FU2021}, \cite{LIU2021}, \cite{Kripa2021}, \cite{Guerar2022} \\
Intellectual Property Rights & \cite{ZEILINGER2018}, \cite{Li2020}, \cite{KONASHEVYCH2020}, \cite{KUDUMAKIS2020} \\
\\
\hline
\end{tabular}
\caption{Literature clustering into main topics based on document content.}
\label{Tab:ScopusClusters}
\end{table}

\subsubsection{Digital Rights Management} \label{Sec:DRM}
This topic includes all documents addressing the use of blockchain technologies for managing the media lifecycle taking into account its copyright. They range from those about media registration to prove ownership to copyright transfer, licensing or controlled access by consumers. All of them explore the use of blockchain technologies to enhance systems that support different parts of this lifecycle. For instance, \cite{CHEN2018} focuses on improving over-the-top (OTT) media services, which offer it directly to viewers using Internet technologies. Similarly, \cite{KUO2019} applies blockchain technologies for access control, though in this case for medical content. Other papers, like \cite{GARBA2021}, address both content distribution but also registration. In contrast, \cite{HOLLAND2017} or \cite{ENGELMANN2018} focus on the exchange of certification and license data, in this case, about 3D models between owner and print service providers. Finally, \cite{GARCIA2021}, in addition to registration, applies blockchain technologies like Non-Fungible Tokens (NFTs) to copyright transfer and licensing.

\subsubsection{Copyright Protection} \label{Sec:CopyrightProtection}
This topic includes papers about blockchain-based mechanisms to improve media protection against piracy or fake content. Most focus on content identification mechanisms to help creators register their content and detect near-duplicates potentially infringing their rights. Likewise, \cite{DOBRE2020} contributes an algorithm that can extract a signature that is resistant to different levels of JPEG compression. In both cases, the hash is stored on the blockchain along with the identification data of the copyright owner. Then, they can use it to detect copies when someone tries to register the same or a similar image, as determined by the algorithm. Other examples are \cite{Stallin2022}, which uses watermarks for copyright protection, or \cite{Igarashi2021}, enabling photos traceability through certified digital cameras.

\subsubsection{Social Media} \label{Sec:SocialMedia}
All papers classified under this topic are those that, among other aspects, place their focus on social media copyright management. For instance, \cite{Kripa2021} also addresses copyright protection using a method of hashing images that is resistant to modification, rotation, and colour alteration. However, unlike papers on the previous topic, it focuses on applying it in the context of social media. Another example is \cite{FU2021}, which also explores copyright protection but in the context of a particular social media platform and its business model. On the other hand, \cite{DASKAL2020} focuses on a regulatory perspective, the creation of a blockchain-enabled network of ombudspersons that help to deal with malicious content like fake news in social media. Finally, also connected with fake content in social media, \cite{GARCIA2019} and \cite{Guerar2022} report about the application of blockchain technologies for the management of verified social media to facilitate its re-use for journalistic purposes.

\subsubsection{Intellectual Property Rights} \label{Sec:IPR}
This topic collects all the papers dealing with the legal aspects of media copyright management and the opportunities offered by blockchain technologies in this context. For instance, \cite{Li2020} proposes to add a remix right implemented using blockchain technologies. This new right would include some elements of compulsory licensing and Creative Commons, allowing remixers to do so without permission but requiring proper attribution and remuneration. Finally, another paper addressing legal issues is the one about applying distributed ledger technologies to real estate, property rights and public registries \cite{KONASHEVYCH2020}. Though the main topic deviates from media copyright, the legal implications analyzed regarding legal identity and privacy are also interesting from the copyright perspective.

\section{Business Review}\label{Sec:BusinessReview}

Complementing the review of media copyright management using blockchain from the academic perspective in Section~\ref{Sec:LiteratureReview}, this section addresses activities in the business sector. The focus is placed on the potential of blockchain technologies to disrupt existing business models and generate new ones. The impact of blockchain in the media industry is even more relevant due to the profound changes that digitization and the Internet have caused.

The issues caused by digitization and the Internet are still there even after the widespread adoption of new business models like streaming. In fact, though streaming might have generated new opportunities for digital service providers, it has made things even worse for other media industry actors, especially creators \cite{Sisario2021}.

Existing initiatives trying to apply blockchain technologies in the media industry are analyzed. First, to better provide an overview of the market, the primary use cases these initiatives try to address are identified. We have considered existing reports about media and blockchain use cases to provide a relevant and diverse set.

The reports under consideration are Deloitte's from 2017 \cite{Deloitte2017}, Protokol's from 2020 \cite{Protokol2020}, JP Morgan's also from 2020 \cite{JPMorgan2020} and The Capital's from 2021 \cite{Shilina2021}. For convenience, the full list of use cases proposed by each report is shown in \textbf{Table~\ref{Tab:ProposedUseCases}}. The table also shows a consolidated set of business use cases we propose, as detailed later in this section. 

Most of the reports consider the media industry in general, though the report from The Capital \cite{Shilina2021} focuses on music. It is also relevant as music is one of the most active media domains and a reference for the others due to its complexity and the wide range of actors involved. All reports consider the whole media value chain, from content creators, including aggregators, platform providers and, when relevant, collecting societies handling royalty payments.

The first impression after analyzing the previous reports is that most of them seem to follow the path established by the oldest one, Deloitte's report from 2017 \cite{Deloitte2017}. The only report that is clearly outside this trend is the most recent one by The Capital \cite{Shilina2021}.

The first use case identified in Deloitte's report is \textit{Use Case 1: New pricing options for paid content}. It focuses on micro-payments as a mechanism to make new pricing opportunities arise. As mentioned, Protokol's and JP Morgan's reports follow the path set by this report and also include micropayments, also as \textit{Use Case 1} in the case of JP Morgan's and as \textit{Use Case 2}, also including usage-based payment models, in Protokol's report. 

On the other hand, the newest report by The Capital does not consider micropayments as a separate use case. As shown in Table~\ref{Tab:ProposedUseCases}, the set of use cases we propose after analyzing the previous reports does not include micropayments. Our view is that this is not a separate use case any longer. Micropayments have not become a primary driver in the media industry and, in any case, they are used in combination with other use cases.

The second use case in Deloitte's report is \textit{Use Case 2: Content bypassing aggregators}. Its focus is mainly on bypassing aggregators from a media marketing perspective, but the use case also includes distributors. A similar use case is also present in Protokol's report, though focusing just on advertising as stated in its title \textit{Use Case 3: Immutable Advertising Engagement Metrics}. Protokol's report includes a separate use case about disintermediation from the distribution perspective. Similarly, both JP Morgan's and The Capital's reports feature use cases about disintermediation mainly from the content distribution perspective.

Additionally, The Capital's report includes one use case related to engagement but as \textit{Monetary incentives for listeners}. Since most reports separate the aggregation and distribution dimensions when talking about disintermediation use cases made possible by blockchain technologies, we propose to consider them as two separate use cases, as shown in Table~\ref{Tab:ProposedUseCases}. The proposed ones are \textit{Use Case 3: Marketing, Fan Engagement and Fundraising} and \textit{Use Case 4: Disintermediated Distribution}.

The third use case proposed in Deloitte's report is \textit{Use Case 3: Distribution of royalty payments}, which is also present in Protokol's and JP Morgan's reports. The Capital proposes more detailed use cases related to copyright management, dealing with specific aspects that allow implementing royalty distribution using blockchain technologies. These are associated with a digital rights database, tokenized rights management and data transparency regarding revenue streams. 

Our proposal is a more general \textit{Use Case 1: Copyright Management}, shown at the top of Table~\ref{Tab:ProposedUseCases}. It goes beyond royalty distribution and includes other aspects required for properly splitting royalties, thus accommodating a broader view of copyright management facilitated by blockchain technologies. This use case includes using smart contracts to carry out royalties’ splits and content registration to associate creators and their content. Moreover, they facilitate linking creations to use conditions that determine how royalties are generated and distributed.

The fourth use case in Deloitte's report is \textit{Use Case 4: Secure and transparent C2C sales}, which is also present in JP Morgan's report. Protokol's report proposes a broader perspective on consumer-to-consumer sales focusing on fraud and piracy prevention. Our view is that all these use cases try to address one of the main weaknesses of digital content from the copyright perspective, which is the lack of the scarcity constraint that drove many business models before the digitization revolution, especially those connected with art and collection. Blockchain technologies make scarcity possible in the digital world, and this is why we propose to consider a more generic use case called \textit{Use Case 2: Digital Content Scarcity}. 

This use case includes the previous use cases by addressing the scarcity issue, but also the related one in the report by The Capital called \textit{New revenue sources for artists}. The latter mostly corresponds to new revenue streams that digital scarcity makes possible, though it also overlaps with our proposed \textit{Use Case 3}, as shown in Table~\ref{Tab:ProposedUseCases}. This overlap is because some aspects regarding the connection with consumers, like fundraising, might also become new sources of revenue for creators.

Finally, Deloitte's report also proposes a fifth use case, \textit{Use Case 5: Consumption of paid content without boundaries}. We have also included this use case in our proposed \textit{Use Case 2: Digital Content Scarcity} because the mechanisms introduced by blockchain technologies regarding scarcity are not constrained, at least technically, by country or regional boundaries. Thus, they can also be used to address this use case.

The result of our review of media and copyright business use cases involving blockchain technologies, consolidating the different reports that we have considered, is the following list of main use cases:

\begin{itemize}
\item {\it Use Case 1: Copyright Management}.
\item {\it Use Case 2: Digital Content Scarcity}.
\item {\it Use Case 3: Marketing, Fan Engagement and Fundraising}.
\item {\it Use Case 4: Disintermediated Distribution}.
\end{itemize}

The summary of how the proposed use cases relate to all the considered ones is summarised in Table~\ref{Tab:ProposedUseCases}, which tries to highlight using ranges of similar colours those use cases with some similarities across the different reports and our proposed set of use cases. In the following subsections, we present each use case and illustrate them through existing initiatives trying to address each using blockchain technologies in the media industry context. The final objective is to have a clearer picture of the domain from the business perspective.

\begin{sidewaystable}[p]
\renewcommand{\arraystretch}{1.25}
\centering
\begin{tabular}{llllll}
\multicolumn{1}{r}{\cellcolor[HTML]{FFFFFF}\textbf{\begin{tabular}[c]{@{}r@{}}\\ Analised Reports \\ \\ Proposed Use Cases\end{tabular}}} & \multicolumn{1}{c}{\textbf{\begin{tabular}[c]{@{}c@{}}Deloitte\\ (2017)\end{tabular}}} & \multicolumn{1}{c}{\textbf{\begin{tabular}[c]{@{}c@{}}Protokol\\ (2020)\end{tabular}}} & \multicolumn{1}{c}{\textbf{\begin{tabular}[c]{@{}c@{}}JP Morgan\\ (2020)\end{tabular}}} & \multicolumn{1}{c}{\textbf{\begin{tabular}[c]{@{}c@{}}The Capital\\ (2021)\end{tabular}}} &  \\
\cellcolor[HTML]{B6D7A8} & \cellcolor[HTML]{B6D7A8} & \cellcolor[HTML]{B6D7A8} & \cellcolor[HTML]{B6D7A8} & \cellcolor[HTML]{D9EAD3}\begin{tabular}[c]{@{}l@{}}A digital rights \\ database\end{tabular} &  \\
\cellcolor[HTML]{B6D7A8} & \cellcolor[HTML]{B6D7A8} & \cellcolor[HTML]{B6D7A8} & \cellcolor[HTML]{B6D7A8} & \cellcolor[HTML]{93C47D}\begin{tabular}[c]{@{}l@{}}Tokenized rights \\ management\end{tabular} &  \\
\multirow{-3}{*}{\cellcolor[HTML]{B6D7A8}\textbf{\begin{tabular}[c]{@{}l@{}}Use Case 1:\\ Copyright \\ Management\end{tabular}}} & \multirow{-3}{*}{\cellcolor[HTML]{B6D7A8}\begin{tabular}[c]{@{}l@{}}Use Case 3:\\ Distribution of \\ royalty payments\end{tabular}} & \multirow{-3}{*}{\cellcolor[HTML]{B6D7A8}\begin{tabular}[c]{@{}l@{}}Use Case 1:\\ Streamlined \\ Royalty Payments\end{tabular}} & \multirow{-3}{*}{\cellcolor[HTML]{B6D7A8}\begin{tabular}[c]{@{}l@{}}Use Case 3:\\ Royalty \\ Distribution\end{tabular}} & \cellcolor[HTML]{6AA84F}\begin{tabular}[c]{@{}l@{}}Complete transparency \\ and data protection\end{tabular} &  \\
\cellcolor[HTML]{FFFFB4} & \cellcolor[HTML]{FFFF00}\begin{tabular}[c]{@{}l@{}}Use Case 4: \\ Secure and transparent \\ C2C sales\end{tabular} & \cellcolor[HTML]{FFFFB4} & \cellcolor[HTML]{FFFF00}\begin{tabular}[c]{@{}l@{}}Use Case 4:\\ C2C Sales\end{tabular} & \cellcolor[HTML]{FFE599} &  \\
\multirow{-2}{*}{\cellcolor[HTML]{FFFFB4}\textbf{\begin{tabular}[c]{@{}l@{}}Use Case 2:\\ Digital Content \\ Scarcity\end{tabular}}} & \cellcolor[HTML]{FCFCD8}\begin{tabular}[c]{@{}l@{}}Use Case 5:\\ Consumption of \\ paid content \\ without boundaries\end{tabular} & \multirow{-2}{*}{\cellcolor[HTML]{FFFFB4}\begin{tabular}[c]{@{}l@{}}Use Case 5:\\ Fraud and Piracy \\ Prevention\end{tabular}} &  & \cellcolor[HTML]{FFE599} &  \\
\cellcolor[HTML]{F4CCCC} & \cellcolor[HTML]{D9D2E9} & \cellcolor[HTML]{F4CCCC} & \cellcolor[HTML]{DAD3E9} & \multirow{-3}{*}{\cellcolor[HTML]{FFE599}\begin{tabular}[c]{@{}l@{}}New revenue sources \\ for artists\end{tabular}} &  \\
\multirow{-2}{*}{\cellcolor[HTML]{F4CCCC}\textbf{\begin{tabular}[c]{@{}l@{}}Use Case 3:\\ Marketing, \\ Fan Engagement\\ and Fundraising \end{tabular}}} & \cellcolor[HTML]{D9D2E9} & \multirow{-2}{*}{\cellcolor[HTML]{F4CCCC}\begin{tabular}[c]{@{}l@{}}Use Case 3:\\ Immutable Advertising\\ Engagement Metrics\end{tabular}} & \cellcolor[HTML]{DAD3E9} & \cellcolor[HTML]{EA9999}\begin{tabular}[c]{@{}l@{}} \\ \\ Monetary incentives \\ for listeners \vspace{5 mm} \end{tabular} &  \\
\cellcolor[HTML]{C9DAF8}\textbf{\begin{tabular}[c]{@{}l@{}}Use Case 4:\\ Disintermediated\\ Distribution\end{tabular}} & \multirow{-3}{*}{\cellcolor[HTML]{D9D2E9}\begin{tabular}[c]{@{}l@{}}Use Case 2:\\ Content bypassing \\ aggregators\end{tabular}} & \cellcolor[HTML]{C9DAF8}\begin{tabular}[c]{@{}l@{}}Use Case 4:\\ Disintermediation\end{tabular} & \multirow{-3}{*}{\cellcolor[HTML]{DAD3E9}\begin{tabular}[c]{@{}l@{}}Use Case 2: \\ Elimination of \\ Content Aggregation\end{tabular}} & \cellcolor[HTML]{C9DAF8}\begin{tabular}[c]{@{}l@{}}The ability to remove \\ middlemen\end{tabular} &  \\
\textbf{(Micropayments)} & \cellcolor[HTML]{FCE5CD}\begin{tabular}[c]{@{}l@{}}Use Case 1:\\ New pricing options\\ for paid content\end{tabular} & \cellcolor[HTML]{FCE5CD}\begin{tabular}[c]{@{}l@{}}Use Case 2:\\ Micropayments \\ and Usage-Based\\ Payment Models\end{tabular} & \cellcolor[HTML]{FCE5CD}\begin{tabular}[c]{@{}l@{}}Use Case 1:\\ Micropayments \\ for Content\end{tabular} &  &  \\
\vspace{5 mm}
\end{tabular}
\caption{Consolidated set of blockchain and media business use cases from existing reports.}
\label{Tab:ProposedUseCases}
\end{sidewaystable}

\subsection{Use Case 1: Copyright Management} \label{Sec:UseCase1}

This use case considers the full copyright life cycle as presented in Section~\ref{Sec:IntroCopyrightBlockchain}. It starts from copyright inception when a creator first manifests a new work into something tangible (a Manifestation). As detailed in the next subsections, there are many initiatives addressing that part of this use case because blockchain technologies facilitate time-stamping those manifestations and linking them to the claimed creator.

Another relevant part of the copyright life cycle considered by initiatives addressing this use case is to track all the actions along the copyright value chain once authorship has been set. This includes consumption by end users or facilitating splitting of royalties’ payments to all the involved actors. The following subsections also illustrate that part of the use case through different business initiatives.

\subsubsection[WIPO Proof]{WIPO Proof\footnote{\url{https://wipoproof.wipo.int/wdts}}} is an example of a business initiative addressing this use case, particularly the first step on the value chain. It is a digital service that provides a time-stamped digital fingerprint of any file, proving its existence at a specific time. These records can be then used as trusted digital evidence. Other similar services are FileProtected\footnote{\url{https://www.fileprotected.com}} or Binded\footnote{\url{https://binded.com}}.

\subsubsection[Kelp Digital]{Kelp Digital\footnote{\url{https://kelp.digital}}} aims to make photography copyrights easy to check and prove by creating verifiable digital statements associated with the image and rendered with it, together with all the associated licenses and copyright transfer. To do so, Kelp Digital first verifies ownership over the physical equipment used to generate the creation. Currently, ownership validation and copyright claims are available only for professional photo equipment, called Proof of Camera \& Lens ownership. The copyright statements and transaction records are stored on Kep's blockchain.

\subsubsection[Unison]{Unison\footnote{\url{https://www.unisonrights.es/en/}}} aims to facilitate managing, collecting and distributing royalties in a simple, fair and efficient way using blockchain technology. External services are used to track the use of music, which is then analyzed to pay creators timely. Unison provides access to a broad, high-quality music catalogue for music users such as TV channels, radio stations, hotels, gyms, or store chains. Users will pay exclusively for the music they use without approximations or estimations. Similar or related initiatives, also focusing on the music industry, are Blòkur\footnote{\url{https://www.blokur.com}} and Verifi Media\footnote{\url{https://www.verifi.media}}.

\subsubsection[Revelator]{Revelator\footnote{\url{https://revelator.com}}} focuses on later steps in the value chain, to ease the management of digital rights and royalties. It can simplify the complex calculation of multi-licensor and multi-territory rights administration. This copyright platform is designed to track and capture the value of digital music for all rights owners in the copyright chain. Revelator uses this information to speed up royalties operations, including splits with collaborators. Other similar initiatives are Vevue \footnote{\url{https://www.vevue.com}} or FilmChain\footnote{\url{https://filmchain.co}}, which focuses on the film and TV industries.

\subsubsection[The Creative Passport]{The Creative Passport\footnote{\url{https://www.creativepassport.net}}} is a verified digital identifier that allows music creators to update, manage and control all information about them and their works. It can push updated profile information into other music services and pull relevant information from them or music representatives. This digital identity aims to become a unique login solution for music services. Moreover, the creator's identity can be verified by linking it to a government identifier or other industry identifiers like IPI, IPN, ISNI.

\subsection{Use Case 2: Digital Content Scarcity} \label{Sec:UseCase2}

This use case includes many topics in the analyzed use case reports, including consumer-to-consumer sales or fraud and piracy prevention. Moreover, part of the use case is about new revenue sources for artists. All the previous have in common benefit from a feature evident in the physical world and traditionally the basis of copyright law. This feature is scarcity, something missing for a long time in the digital space due to the ease of copying the same bits repeatedly.

Though digital copies are a feature in many senses, which eases scale economies on top of the Internet, it introduces issues like piracy or pressing down the value of content in digital form. Cryptographic mechanisms can be used on top of blockchains to introduce scarcity of digital assets, using unique tokens that can be owned, traded and verified to prevent piracy.

The solution is Non-Fungible Tokens (NFT). Unlike fungible tokens that are interchangeable, like cryptocurrencies or fiat money, they present some unique properties that make them non-interchangeable, i.e. non-fungible. This uniqueness can be tied to digital content like a song or a picture, making it ownable and scarce. 

The only weak point is the link between the NFT and the digital content, especially if it points to a file in centralized storage. Alternatively, to strengthen this link, digital content can be stored on-chain, usually just if it means a small amount of data or code that generated the content, or off-chain but using decentralized storage.

NFTs are also being used to represent ownership of many other assets, from stocks to houses. In these cases, mechanisms are also required to provide trustful ties between the token and the asset. It is usually helpful to think about NFTs as some kind of “receipt”. You own a piece of digital crypto art by proving control of the “receipt” NFT, but the content file for the work might be replicated many times across the Internet. 

That piece might be even a meme, copied thousands of times across social media. However, you can prove that you hold the unique NFT linked to its ownership. At this point, the real issue is if the person who mints the token, from the point of view of copyright law, holds the copyright supposedly transferred through NFT ownership. It is necessary to combine NFTs with systems capable of managing copyright, like those described for \textit{Use Case 1} in the previous section. Thus, it becomes essential to have a way to prove authorship and enable tracing it from the NFT.

\subsubsection[Valuables by Cent]{Valuables by Cent\footnote{\url{https://v.cent.co }}} is one of the easiest ways to create NFTs. It allows minting an NFT for any publicly available tweet. It is also possible to buy tweets from other users, which should be publicly available. The NFT metadata pointing to the referenced tweet is signed using the creator's private key, so we can say that they autograph the NFTs. The process is integrated into the social network, Twitter in this case, as the media is initially available there, and the NFT metadata points to the corresponding tweet. Even if the original tweet is erased by its creator, the metadata included in the NFT will remain as it is available on-chain. Moreover, a screenshot of the tweet is also stored in Cent’s servers. However, it is important to note that just the image corresponding to the tweet is stored, not the full content if the tweet includes an animated GIF or a video. Additionally, if Cent’s servers go down or the service is discontinued, that screenshot will be lost as it is just available in centralized storage. 

\subsubsection[Zora]{Zora\footnote{\url{https://zora.co}}} is an NFTs marketplace that allows creators to define a configurable percentage of future sales of their NFTs. This percentage of sales beyond the first one implements a mechanism like royalties, though it is a proprietary solution and only works for sales on the Zora marketplace.
Zora is also developing the Catalog platform on top of the Zora Protocol, allowing artists to mint their music as one-of-one NFTs, i.e. artists can just press one edition of their music works. Songs are free to listen to everyone and individually ownable by collectors. In addition to the royalties-like feature provided by the Zora Protocol, the plans include that Catalog also supports revenue splits for collaborators. 

\subsubsection[HENI NFT]{HENI NFT\footnote{\url{https://nft.heni.com}}} provides a NFT marketplace for digital art. Through limited editions, HENI shows how blockchain technologies are used to introduce scarcity into digital art and provide new revenue streams for digital artists.

\subsection{Use Case 3: Marketing, Fan Engagement and Fundraising}  \label{Sec:UseCase3}

This use case includes all mechanisms to manage and improve the communication between creators and consumers, and it aims to create a much more direct connection between them. Nowadays, the emergence of aggregators or streaming services makes creators unaware of how their creations are being consumed.

A clear example of this is streaming data. All happens through big platforms like Spotify, which have access to all the aggregated data while it is hard for the artist to get feedback beyond overviews and no way to get it promptly. Blockchain technologies might help to build these channels for direct communication with fans. And this goes beyond usage information, which might also be used for royalties’ payments as described for \textit{Use Case 1} in Section~\ref{Sec:UseCase1}.

New opportunities include using tokens for fan engagement, i.e. a kind of “Proof of Fandom”. These tokens can provide additional incentives like ticket discounts or verifiable merchandise. Or they can be accompanied by loyalty badges or reward tokens.

Another interesting approach is to engage fans to play the role of “Curators” of different kinds of media registries using incentivized strategies like Token Curated Registries \cite{Hassanien2021}. For instance, to offer rewards to fans for curating personalized playlists. Or reward fans with a native token for contributing to a database of artists, venues or events.

Finally, blockchain facilitates artists going into fundraising campaigns that help to align artists' and fans' interests. This kind of crowdfunding helps creators get more independent from centralized sources of funds and makes it possible for consumers to invest and trade in the creators they like.

\subsubsection[DAOrecords]{DAOrecords\footnote{\url{https://www.daorecords.org}}} is both a record label and a platform to connect musicians and artists to their fans using blockchain technologies. Artists have complete control over their music and their relationship with their fans and community. Additionally, DAOrecords is experimenting with the Crypto Art space, minting on-chain Audio NFTs and hosting The Popup, an art and music event series in the Cryptovoxels metaverse.

\subsubsection[RAC]{RAC\footnote{\url{https://rac.fm}}} is the first Portuguese artist to win a Grammy and one of the first musicians to sell his music using blockchain technologies in 2017. The album purchase is represented on-chain by the EGO token. RAC has recently rewarded his fans, for instance, those holding an EGO token, with his community token called RAC. A RAC holder can access a private discord server or receive exclusive early access to future merchandise. Future plans include tokenized advertisement space on RAC’s Twitch channel, discounts on merchandise or access to unique crypto-artwork.

\subsubsection[Steemit]{Steemit\footnote{\url{https://steemit.com}}} stores content in an immutable blockchain and rewards users for their contributions with a digital token called STEEM. The Steem blockchain mints new STEEM tokens every day and adds them to a community's rewards pool. These tokens are then awarded to users for their contributions, based on their content's votes. Users who hold more tokens in their account will decide where a larger portion of the rewards pool goes. Up to 50\% of a post's payout is awarded to curators, who upvoted the post first, as a reward for discovering relevant content. The other 50\% is awarded to the author. A similar initiative is Cent\footnote{\url{https://beta.cent.co}}.

\subsubsection[YellowHeart]{YellowHeart\footnote{\url{https://yh.io}}} is a blockchain-powered ticketing company whose mission is to eradicate scalping and bad players in the secondary ticketing market, thus putting the power back into the hands of fans and artists. Moreover, they consider the rest of the ticketing ecosystem by rewarding venue promoters and the resellers themselves. YellowHeart uses blockchain technologies, particularly smart contracts, to set concert ticket rules. How many seats there are and how much do they cost? These rules include what they can be resold for, how many times, and even where to resell money goes. For instance, split among artists and promoters or entirely to charity.

\subsection{Use Case 4: Disintermediated Distribution} \label{Sec:UseCase4}

This use case includes all disintermediation actions facilitated by blockchain technologies that allow creators to distribute their content to consumers without intermediaries. These intermediaries control distribution channels, including music streaming platforms, and thus can easily influence what content is consumed.

Efforts to change this situation include different kinds of utility tokens that provide access to alternative and decentralized content platforms, for instance, bandwidth tokens for music consumers to compensate creators. Consumers give them part of their bandwidth to reach more consumers without having to rely on other centralized distribution channels.

\subsubsection[Livepeer]{Livepeer\footnote{\url{https://livepeer.org}}} is looking to build a decentralised infrastructure of video transcoding. Developers can use it for adding live video to their projects using the Livepeer public network. Video miners run a Livepeer node and transcode video on their GPUs for token rewards. The network is secured by token holders, who help improve and secure the Livepeer network by acquiring and staking the reward token on video miners. They also are rewarded if they stake in productive video miners. Other examples of initiatives about using blockchain to facilitate media distribution are Audius\footnote{\url{https://audius.org}} or D.Tube\footnote{\url{https://d.tube}}.

\subsubsection[Resonate]{Resonate\footnote{\url{https://resonate.is}}} is a music streaming cooperative that allows listeners to "pay as you a stream" until they own the song. It’s a new listening model called “stream to own.” Only pay for what you play, making a seamless transition from casual listening into becoming a dedicated fan. Resonate is a cooperative owned by the musicians, indie labels, fans and workers that build it.

\subsubsection[Contentos]{Contentos\footnote{\url{https://www.contentos.io}}} uses a dedicated blockchain to build a decentralized digital content community that allows content to be freely produced, distributed, rewarded and traded while protecting author rights. With a decentralized revenue system, the value of creation is open, transparent, and returns rewards directly to users. Through rewards, users are encouraged to share and promote content to the right audience. Users are responsible for their credit score, calculated based on every contribution they make. Blockchain technology enables copyright authentication and transactions to be trackable. Joystream\footnote{\url{https://www.joystream.org}} makes a similar proposal, materialized as a decentralized platform for streaming and sharing video content.

\subsection{Evaluation} \label{Sec:Evaluation}

To evaluate the completeness of the proposed use cases, we have considered 31 scenarios applying blockchain to the media industry. These scenarios were collected by Jack Spallone. He was involved from 2017 to 2020 in Ujo Music\footnote{\url{http://ujomusic.com}}, one of the first initiatives applying blockchain to the music industry and currently is currently Head of Crypto at HIFI Labs. Jack describes them in a set of tweets\footnote{\url{https://twitter.com/JackSpallone/status/1377423491746557955}}. Each use case has been classified into one of the proposed use cases:

\begin{itemize}
\item {\it Use Case 1: Copyright Management}: Payment Splits, Right Registry (TCR), Artist Identity, Per-stream Payments, Usage and Reporting, On-Chain Licensing for Off-Chain Rights, NFTs as Synch Licenses.
\item {\it Use Case 2: Digital Content Scarcity}: Non-transferable Token as Access, Bonding Curves to Price Music, NFT as License, Scarce Sounds Marketplace, Scarce Music Releases, 1 of 1 Digital Records.
\item {\it Use Case 3: Marketing, Fan Engagement and Fundraising}: Tipping, NFT as Recording Advance, NFTs as Proof-of-Patronage, Tickets w/ Secondary Market Price Capture, Music Chart Curation with Token Rewards, Non-Copyright Record Deals, Music Crypto Community Token, Community Token Fan Club, Streaming Payment Advances using DeFi, Retroactive Airdrop of Social Tokens, Social Token Community Fund, Stake Social Tokens to Earn Song NFTs, Physical goods redeemed by tokens sold on a bonding curve.
\item {\it Use Case 4: Disintermediated Distribution}: Programmatic Licensing, Streaming Co-op, Publishing DAO, Label DAO, Market-making Distribution Models.
\end{itemize}

As can be observed, the proposed used cases are complete as they accommodated all of the scenarios.

\section{Discussion} \label{Sec:Discussion}

From the literature analysis about media copyright management using blockchain, we highlight that the number of publications indexed by Scopus or Web of Science, including journals and conferences, is relatively low compared to other topics. Just 31 papers have been retrieved.

From an overview of this literature, analyzing aspects like publications per year, subject area or type, we can conclude that this is a very young and still immature area from an academic perspective. In addition to the small number of documents available, their time span is very narrow, starting in 2017 and with almost half of the papers originating in 2020 plus a decline to just six documents in 2021.

Considering their subject areas, Computer Science accumulates one-third of the documents and, together with Engineering and Mathematics, they account for more than 60\%. On the other hand, Social Sciences or Business and Management represent 9\% and 5.1\% respectively. It is also important to note the absence of published reviews when considering the types of documents.

Beyond this overview, this academic literature has also been analyzed in detail. First, we used an automated approach to cluster the documents based on their content. As a result of this analysis, we identified the following four main topics:

\begin{itemize}
    \item \textit{Digital Rights Management}: this topic clusters all the documents focusing on the management of media copyright lifecycle using blockchain technologies. From registration to licensing or controlled consumption.
    \item \textit{Copyright Protection}: the documents classified under this topic propose blockchain-based mechanisms for media protection to fight copyright infringement.
    \item \textit{Social Media}: though papers under this topic also address copyright management and protection issues, their focus is on the specificities of social media.
    \item \textit{Intellectual Property Rights}: this topic includes the documents dealing with the legal aspects of media copyright management, focusing on the opportunities that blockchain technologies bring from a legal standpoint.
\end{itemize}

We contextualized each topic by providing details about some of the corresponding documents, as detailed in Section~\ref{Sec:LiteratureAnalysis}. The complete list of all the documents in each topic is presented in Table~\ref{Tab:ScopusClusters}.

The academic review of media copyright management using blockchain technologies has been complemented from the business perspective. The starting point has been analyzing existing reports and identifying the most relevant use cases to apply blockchain to the media industry. Four relevant reports have been identified, by Deloitte \cite{Deloitte2017}, Protokol \cite{Protokol2020}, JP Morgan \cite{JPMorgan2020} and The Capital \cite{Shilina2021}. 

The analysis has consolidated all the use cases proposed by these reports into four: \textit{Use Case 1}: Copyright Management, \textit{Use Case 2}: Digital Content Scarcity, \textit{Use Case 3}: Marketing, Fan Engagement and Fundraising and \textit{Use Case 4}: Disintermediated Distribution. Table~\ref{Tab:ProposedUseCases} provides an overview of the consolidation process.

To evaluate the completeness of the proposed use cases, we have successfully classified 31 scenarios applying blockchain to the media industry into one of them, as detailed in Section~\ref{Sec:Evaluation}. The evaluation shows that the proposed use cases are complete as they accommodated all of the scenarios. 

Detailed descriptions of each use case are provided in Section~\ref{Sec:BusinessReview}, together with 14 representative examples of business initiatives and 11 similar additional ones. Altogether, 25 initiatives illustrate the scope of each use case and show a very active business ecosystem, in many cases far beyond the state of the art in academic literature.

For instance, the initiatives highlighted for \textit{Use Case 1: Copyright Management} implement solutions that are beyond those depicted in the papers related to the topic \textit{Digital Rights Management}, which in all cases are at most just proofs of concept. It is fair to note that there has been a lot of funding from venture capital, Initial Coin Offerings (ICOs), and other crowdfunding mechanisms for blockchain-related initiatives. For instance, more than USD 31 billion was raised via ICOs (03/2020) since 2016 \cite{Schuckes2021}. This economic inflow seems to have boosted the blockchain industry beyond the state of the art in academia.

Overall, most of the literature is related to \textit{Use Case 1: Copyright Management}, which is related to the main literature topics \textit{Digital Rights Management} and \textit{Copyright Protection}. On the other hand, little literature addresses the other use cases, especially Use Case 3 and 4, which are the most related to new business models emerging from applying blockchain technologies to the media industry. Combined, these facts highlight the importance of taking both the academic and business perspective when reviewing emerging and very business-oriented domains like blockchain and media.

\section{Conclusions} \label{Sec:Conclusions}

Based on the previous analysis and discussion about the situation regarding the use of blockchain technologies for media copyright management, from both academic and business perspectives, it is possible to address the research questions highlighted in the introduction as detailed next.

\textbf{RQ1: is the application of blockchain technologies to media copyright management a mature academic research area?}

Many results point in the direction of a lack of maturity. First of all, the small number of publications in quality journals and conferences indexed by Scopus or Web of Science. Additionally, there is a downtrend since 2020 in the number of publications and most of them are in conferences, less than a third of them in journals and a complete absence of review papers on this topic.

\textbf{RQ2: which are the main areas of academic research dealing with media copyright management using blockchain?}

The analysis of the retrieved literature dealing with blockchain for media copyright management highlights four main topics under which it can be classified. They are \textit{Digital Rights Management}, \textit{Copyright Protection}, \textit{Social Media} and \textit{Intellectual Property Rights}.

\textbf{RQ3: which are the main business use cases for media copyright management using blockchain?}

The main use cases that have been identified are \textit{Use Case 1: Copyright Management}, \textit{Use Case 2: Digital Content Scarcity}, \textit{Use Case 3: Marketing, Fan Engagement and Fundraising} and \textit{Use Case 4: Disintermediated Distribution}.

\textbf{RQ4: where do most "blockchain for media copyright management" initiatives originate, academia or industry?}

The bigger number of initiatives emerging from industry compared to those from the academic world shows that the former is much more active in this topic. It has been possible to identify more than 25 business initiatives addressing all business use cases. Moreover, it has been possible to also classify into the identified use cases 31 scenarios applying blockchain to the media industry.

As a final takeaway from the previous conclusions, we think the results make clear that any kind of research work on this particular topic coming from academia has to pay special attention to what is being done in the industry. Consequently, to keep our review on the use of blockchain technologies for media copyright management updated, our future plans include to continue monitoring academic publications while also analysing grey literature, like white papers or business reports, using the business use cases identified as the analysis framework.

\section*{Funding}
Supported by project ONTOCHAIN, which has received funding from the European Union’s Horizon 2020 research and innovation programme under grant agreement No 957338.

\bibliographystyle{ACM-Reference-Format}
\bibliography{bibliography}

\end{document}